\documentclass[11pt,letterpaper]{article}
\usepackage{amsmath,amssymb,amsfonts}
\usepackage{graphicx,wrapfig}
\usepackage[usenames,dvipsnames]{xcolor}
\usepackage{skak}
\usepackage{caption}
\usepackage[affil-it]{authblk}

\graphicspath{{./figures/}}
\numberwithin{equation}{section}
\allowdisplaybreaks

\newcommand{\be}{\begin{equation}}
\newcommand{\ee}{\end{equation}}
\newcommand{\ba}{\begin{aligned}}
\newcommand{\ea}{\end{aligned}}
\newcommand{\ben}{\begin{eqnarray}\displaystyle}
\newcommand{\een}{\end{eqnarray}}

\newcommand{\cN}{\mathcal{N}}

\setboardfontsize{8}

\begin{document}

\begin{titlepage}

\title{Bootstrapping the $\cN=1$ SCFT in three dimensions.}

\author{Denis Bashkirov
  \thanks{Electronic address: \texttt{dbashkirov@perimeterinstitute.ca}}}
\affil{Perimeter Institute for Theoretical Physics\\
Waterloo, Ontario, ON N2L 2Y5, Canada}

\maketitle

\abstract{We suggest a way to implement conformal bootstrap program for the case of the ${\cal N}=1$ SCFT in three dimensions using the previous analysis of the Ising model in \cite{CB}. We find approximate values for the conformal dimensions of several operators and the central charge $C_T$, the coefficient in the two-point function of the stress-tensor. Boostrapping this particular (minimal) SCFT is of special interest as it was suggested in \cite{CM} that it may realize supersymmetry in 2+1 dimensions in experiment. The values are in a good agreement with the previous estimate in \cite{CM}.}

\end{titlepage}

 In this note we offer a way to apply the three-dimensional conformal boostrap program \cite{CB,CB3} to the minimal ${\cal N}=1$ SCFT in three dimensions. The natural candidate for this model is the infrared fixed point of the UV Lagrangian theory whose superfield is the real scalar superfield $\Sigma$ containing a real scalar $\sigma$ and a Majorana fermion $\psi$. The interactions are the Yukawa interaction $\sigma\psi_\alpha\psi_\beta\epsilon^{\alpha\beta}$ and its superpartner $\sigma^4$ which come from the real superpotential $W=\Sigma^3$, so that the interaction terms are $V=W|_{\theta^2}$. This theory has recently been the subject of a search of supersymmetric theories in condensed matter systems \cite{CM}.

 In this real scalar supermultiplet $\Sigma$ there are only two conformal primaries: the superconformal primary $\sigma$ and another scalar $\epsilon$ which lives on the second level and is $\epsilon=\epsilon^{\alpha\beta}[Q_\alpha,Q_\beta]\sigma$, or in components (omitting the fermion)
\begin{align}
\Sigma=\sigma+\epsilon^{\alpha\beta}\theta_\alpha\theta_\beta\epsilon+...
\end{align}

The scalar $\epsilon$ is the $F-$term giving $F=\sigma^2$ on the $F-$ term equation. Furthermore, it is a conformal primary without any mixing with total derivatives as there are no total derivatives to mix with.

Now consider OPE of a two scalars $\sigma(x)\sigma(0)$ from a real scalar multiplet. Schematically, it is
\begin{align}
\sigma\sigma=1+\sigma^2+...,
\end{align}

where $\sigma^2$ is the composite operator. It is a conformal primary operator in a real scalar multiplet ${\cal O}$,\footnote{We are going to identify this supermultiplet with $\Sigma$ itself momentarily.} so it comes together with another conformal primary scalar. Let us denote the second primary as $\varepsilon$. Thus, in terms of conformal primaries, schematically,

\begin{align}
\sigma\sigma=1+\lambda(a\sigma^2+b\varepsilon)+...,
\end{align}
where the relative values of $a$ and $b$ are fixed by ${\cal N}=1$ superconformal symmetry.

To find this relation we may turn to the superfield formalizm \cite{Park}.

We need to consider the three-point function
\begin{align}
<{\cal O}\Sigma\Sigma>
\end{align}
 From \cite{Park} equation (5.28) we have an ansatz
\begin{align}
<{\cal O}\Sigma\Sigma>=\frac{H(X_1,\Theta_1)}{(x_{12}^2+\frac14\theta_{12}^4)^{\eta_2}(x_{13}^2+\frac14\theta_{13}^4)^{\eta_3}}
\end{align}

 Here $(X_1,\Theta_1)$ are some complicated functions of the supercoordinates. We will not give their explicit form (see \cite{Park}).

$H$ must be a scalar and be homogeneous with weight $(\eta_2+\eta_3-\eta_1=2\Delta_\sigma-\Delta_\epsilon)$.

As in 4d \cite{CB2}, setting $\theta_2=\theta_3=0$ and $\theta_1=\theta$ we get the relation between $<\sigma\sigma\epsilon>$ and $<\sigma\sigma\epsilon'>$.

It is straightforward to expand this expression in $\theta$. There are two such expressions. We do not give expressions for them as this will not be important. The result is that one of the two three-point function is zero. It is easy to understand the reason behind this.

That fact that only one of the operators $\sigma$, $\varepsilon$ appears in the OPE $\sigma\sigma$ is a reflection of the parity properties of the fields.\footnote{We are greatful to Davide Gaiotto for pointing this out to us.} Indeed, the Yukawa term $\sigma\psi^\alpha\psi^\beta\epsilon_{\alpha\beta}$ of the UV Lagrangian implies that the parity is preserved with $\sigma$ being a pseudoscalar. This has the following consequences.

The field $\sigma$ is actually $\varepsilon$ and itself cannot appear in the OPE $\sigma\sigma$, so we can write schematically $\sigma\sigma=1+\sigma^2+...$, where $\sigma^2$ is the composite operator. Furthermore, we can include the next scalar operator $\epsilon'$ (do not confuse it with $\varepsilon=\sigma$!), another primary operator, possibly, for example, $[\sigma^4]$ or $[\sigma^6]$ which we discuss later on, so that $\sigma\sigma=1+\sigma^2+\epsilon'+...$. The structure of this OPE is just like that in the Ising model if we denote $\epsilon=\sigma^2$ and $\epsilon'=\epsilon'$. Of course, the conformal dimensions of these operators are not those in the Ising model.

To justify the identification of $\varepsilon$ with $\sigma$, return to the UV Lagrangian for a moment. The $F-$term equation sets the $F-$component of the real $\Sigma$ supermultiplet to $\sigma^2$, so that
\begin{align}
\Sigma=\sigma+\theta\psi+\theta^2\sigma^2.
\end{align}

This immediately gives us a relation between conformal dimensions of $\sigma$ and $\sigma^2$: $\Delta_{\sigma^2}=\Delta_\sigma+1$. Just as only one of the two conformal primaries of the $\Sigma$-supermultiplet appeared in the $\sigma\sigma$ OPE, only one of the two primaries in the real supermultiplet ${\cal E}$ containing $\epsilon'$ will appear in the same OPE, the one which is a scalar, not a pseudoscalar. We denoted it as $\epsilon'$. This conformal primary may be a superconformal primary or maybe its superdescendant.  This is not important for our analysis.

So at this level of the OPE the difference between the CFT and SCFT data is not seen. The operators on the right hand side come in conformal multiplets, not in superconformal multiplets! Of course, there will be difference in OPE coefficients and conformal dimensions of operators on the right hand side, but given the ignorance about them we do not know the difference. Naively then it seems that there is an obstacle in bootstrapping the superconformal model since we cannot distinguish it from a CFT model, more precisely, the Ising model. However, it is actually helpful, as we will be able to use the numerics for the Ising model and take into account supersymmetry through relations between conformal dimensions of operators in the same superconformal multiplet. Furthermore, the part of numerics for the Ising model that we borrow is universal -- it does not depend on the information about the operators that appear on the RHS of the $\sigma\sigma$ OPE except the very first one, $\sigma^2$. This is a consequence of the analysis of \cite{CB} illustrated in the figures 3 through 5.

Forgetting for a moment about the constraint $\Delta_{\sigma^2}=\Delta_{\sigma}+1$, the steps of implementing the Conformal bootstrap to the four-point function $<\sigma\sigma\sigma\sigma>$ are identical to the Ising model case \cite{CB,CB3} as long as we do not take into account the known bounds on $\epsilon'$ and other operators in the Ising model. This means that we get the same curve for the upper bound on $\Delta_{\sigma^2}$ as a function of $\Delta_{\sigma}$ as the curve for the $\Delta_\epsilon$ as a function of $\Delta_\sigma$ in the Ising model.

In the following we heavily use the numerical results and graphics for the bootstrapped Ising model analyzed in \cite{CB}. Namely, we use figures 3-5 and figure 11.

Now to distinguish our ${\cal N}=1$ minimal SCFT from the minimal CFT which is the IR limit of the Ising model, we take into account the constraint $\Delta_{\sigma^2}=\Delta_\sigma+1$. This line intersect the allowed region for the $(\Delta_\sigma,\Delta_{\sigma^2})$ in figures 3-5 in \cite{CB} and gives a lower bound on the conformal dimension of the $\sigma$ which is approximately $\Delta_\sigma(min)\approx 0.565$ which is, of course, above the conformal dimension of $\sigma$ in the Ising model. We note that this bound does not change noticeably in the figures 3 through 5. Of course, this bound can be improved, that is increased to some extent when imposing some constraints on the conformal dimension of other operators in the $\sigma\sigma$ OPE, for example, $\epsilon'$ and improving the numerics. However, as the analysis of the Ising model suggests, there will be only a very weak dependence on these additional parameters, as the upper part of the curve for the Ising model (the bound curve above the Ising point) is essentially unmodified as is seen in \cite{CB} in figures 3 through 5.

So we know that $\Delta_\sigma\ge 0.565$. Now we make use of the following empirical observation \cite{CB,CB3}: in all known minimal models the central charge $C_T$ defined as the coefficient in the two-point function of the stress-tensor attains its minimal value as a function of $\Delta_\sigma$ allowed by a lower bound\footnote{We would like to thank David Simmons-Duffin for informing us about this fact.}. See figure 11 in \cite{CB}. It was again observed in \cite{CB} for the Ising model that the curve for this bound changed insignificantly under changes of conformal dimensions of other operators in the theory. For example, for $\Delta_\epsilon$ allowed to vary from the unitary bound $1/2$ to the $\Delta_{max}(\Delta\sigma)$ the curve essentially preserved its form and shifted a little to the right. In particular, when we restrict the values of $\Delta_\epsilon=\Delta_{\sigma^2}$ to the supersymmetric ray, the minimum of the function $C_T(\Delta_\sigma)$ will be attained at the minimal allowed $\Delta_\sigma\approx 0.565$. The minimal value is approximately $C_T(0.565)\approx 1.13$ which is, of course, greater than for the Ising Model. Note also, that the conformal dimension of $\sigma$ is in a good agreement with the value $\Delta_\sigma=4/7=0.571$ presented in \cite{CM}.

\section*{Summary}

We should stress that the approximate values $\Delta_{\sigma}\approx 0.565$, $\Delta_{\sigma^2}\approx 1.565$, $\Delta_{\psi}=\Delta_{\sigma}+1/2\approx 1.065$ and $C_T\approx 1.13$ were obtained assuming that the empirical principle of the minimality of the central charge $C_T$ holds for the minimal ${\cal N}=1$ SCFT which we conjectured to be the infrared fixed point of the Lagrangian theory of a real supermultiplet with the real superpotential $W=\Sigma^3$. We are not aware of any fundamental argument in favor of the universal validity of the $C_T-$minimization principle. Nevertheless, it seems to be a natural guess given its validity for all known minimal (in the sense of bootstrap) models in three dimensions and may be the answer to the question ``what quantity is minimal in minimal models in three dimensions?". Furthermore, a good agreement with the estimate in \cite{CM} is evidence in favor of this assumption in the present model. The lower bound $\Delta_\sigma\ge 0.565$ is independent of this assumption and must necessarily hold.

We want to stress again that due to the identical forms of the $\sigma\sigma$ OPE for the first couple of operators on the right hand side for the Ising model and the $\cN=1$ SCFT we could borrow the analysis of the four-point function $<\sigma\sigma\sigma\sigma>$ from the Ising model due to a very weak dependence on the higher-dimension operators for the quantities of interest. These operators are among other things that know the difference between the two theories. The distinction was implemented by imposing the constraint $\Delta_\epsilon=\Delta_\sigma+1$, which is a consequence of ${\cal N}=1$ supersymmetry. One could say that at the level of the four-point function $<\sigma\sigma\sigma\sigma>$ the distinction by additional constraint $\Delta_\epsilon=\Delta_\sigma+1$ is 'relevant' while by higher dimension operators are 'irrelevant' (much weaker) as long as $\Delta_\sigma$ is concerned. The weakness of dependence of the 'upper' part of the curve $\Delta_\epsilon (\Delta_\sigma)$ where it is intersected by the line $\Delta_\epsilon=\Delta_\sigma+1$ when $\Delta_{\epsilon'}$ was varied is demonstrated in figure 5 of \cite{CB}. This is what validates our approach.

\section{Acknowledgements}
The author wishes to thank David Simmons-Duffin and especially Davide Gaiotto for very useful conversations.
This research is supported by the Perimeter Institute for Theoretical Physics. Research at the Perimeter Institute is supported by the Government of Canada through Industry Canada and by Province of Ontario through the Ministry of Economic Development and Innovation.


\begin{thebibliography}{99}


\bibitem{CB} Sheer El-Showk, Miguel F. Paulos, David Poland, Slava Rychkov, David Simmons-Duffin, Alessandro Vichi, ``Solving the 3D Ising Model with the Conformal Bootstrap," [arXiv:1203.6064[hep-th]].

\bibitem {CM} Ashvin Vishwanath, ``Emergent Supersymmetry on the Surface of a Topological Phase," string/CMT group meeting at Perimter Institute, August 2nd, 2013;

Tarun Grover, D. N. Sheng, Ashvin Vishwanath, ``Emergent Space-time Supersymmetry at the Boundary of a Topological Phase," [arXiv:1301.7449]

\bibitem{CB2} David Poland, Slava Rychkov, David Simmons-Duffin, ``Bounds on 4D Conformal and Superconformal Field Theories," {\bf JHEP 1105} (2011) 017  [arXiv:1009.2087[hep-th]].

\bibitem{CB3} Filip Kos, David Poland, David Simmons-Duffin, ``Bootstrapping the O(N) Vector Models,"  [arXiv:1307.6856[hep-th]].

\bibitem{Park} Jeong-Hyuck Park, ``Superconformal Symmetry in Three-dimensions ,'' {\bf J.Math.Phys. 41} (2000) 7129-7161  [arXiv:9910199 [hep-th]].


\end{thebibliography}
\end{document}